\documentstyle[preprint,aps]{revtex}
\begin{document}
\draft
\title{\bf The geometry of RG flows in theory space }
\author{Sayan Kar \thanks{Electronic Address :
sayan@phy.iitkgp.ernet.in//sayan@cts.iitkgp.ernet.in}}
\address{Department of Physics and Centre for Theoretical Studies,\\
Indian Institute of Technology, Kharagpur 721 302, INDIA}
%\twocolumn[
\maketitle
\widetext
\parshape=1 0.75in 5.5in
\begin{abstract}
Renormalisation Group (RG) flows in theory space (the space of
couplings)
are generated
by a vector field -- the $\beta$ function.
Using
a specific metric ansatz
in theory space and certain
methods employed largely in the context of
General Relativity, we examine
the nature of the expansion, shear and rotation of
geodesic RG flows. The expansion turns out to be a negative quantity
inversely related to the norm of the $\beta$ function. This implies
the focusing of the flows towards the fixed points of a given field theory.
The evolution equation for the expansion
along geodesic RG flows is written down and analysed.
We illustrate the results  for a scalar field theory with a $j\phi$ coupling
 and
pointers to other
areas are briefly mentioned.
\end{abstract}

%PACS number(s) :
%\pacs{}
%]

%\narrowtext
\newpage
\vspace{.2in}

In the theory of quantised fields, the notion of the Renormalisation
Group (RG) plays a central role in understanding certain non--perturbative
issues. It is a well-known fact that an
ambiguity arises in the choice of the infinite part of a regularised
Feynman amplitude. This necessitates the choice of a renormalisation scheme,
which,
naturally, implies the existence of a scale or a renormalisation point.
The requirement
that physical amplitudes, say the one-particle irreducible amplitudes
, are independent of a choice of scale leads to the concept
of the Renormalisation Group. The RG transformations
are finite renormalisations and the RG equation, essentially,
follows from the scale invariance of the physical Green functions.

The fact that the RG equation can be written as a {\em geometric
statement} was first noted by Lassig in {\cite{lassig}} and
later by Dolan {\cite{dolan1}}, {\cite{dolan2}}. The idea there, was to
visualise the $\beta$--function as a vector field in the space
of couplings (this is the space we refer to as `theory space'). Theory
space, constructed with the couplings as the coordinates can be
thought of as a manifold in its own right. Each point on such a
manifold defines a set of values for the various couplings which
appear in a given Lagrangian. A curve in theory space, therefore,
represents the flow of couplings. Finite renormalisation
generates such a flow and thereby, gives rise
to RG trajectories which have the $\beta$--function as the tangent
vector at each point.
The RG equation, in its theory space incarnation, can be expressed
as a Lie--derivative of the
n--point function with respect to this vector field.
The change in physical amplitudes due to a diffeomorphism of
$R^{D}$ (scale transformations of the spatial variables), at fixed
couplings, generated by the
dilatation generator $\bf D$ is equivalent to the change in amplitudes
due to a diffeomorphism of theory space, generated by the $\beta$--function,
at fixed spatial positions.
This fact lies at the heart of the geometric form of the RG equation.
As is well known, the definition of the Lie--derivative does not
require the concept of a metric or a connection.
However, we do need such structures
in order to define and analyse a geometry. The natural question
which one therefore has to address is --
how do we define a metric in theory space ?

There have been proposals for such a metric.
One of them is related to the spatial integral of a two--point
function {\cite{oconnor}},
which, in general, could, as well be, a two-point function of
composite operators.
We also need to impose the crucial fact that spatial points are well separated.
This enables us to avoid the possible divergences of amplitudes which
always come up in quantum
field theory.
The above proposal is akin to that
due to Zamolodchikov{\cite{zamolo}}, which he and later authors
utilised in order to prove the $c$-theorem for two--dimensional
field theories.

Given a definition of a metric we can now investigate the {\em geometry}
of theory space through the ensuing connection and
curvature. This has been worked out for several field theoretic
models such as scalar field theories with a $j\phi$ term {\cite{dolan3}},
$\lambda \phi^{4}$ theory, the one dimensional Ising model {\cite{dolan3}},
the O(N)
model (for large N) {\cite{dolan4}} and $N=2$ supersymmetric Yang--Mills
theory {\cite{dolan5}}. The location of zeros of the $\beta$--function
which correspond
to fixed points for the flows, (and also imply conformally invariant
field theories) coincide, rather surprisingly, with points of
diverging Ricci curvature
in theory space.

The fact that the $\beta$
function is a vector field in theory space can
improve our understanding of the nature of RG flows
in a novel way, if we employ the techniques and results
primarily used in proving the singularity theorems in General
Relativity (GR). This is what we aim to do in this
article.  We shall also point out that there are some worthwhile
directions which may be pursued in greater depth, in future.

Firstly, let us recall the analysis from Riemannian geometry and
GR which we will use
extensively. This is based on the fact that the covariant derivative of
the
velocity field in a Riemannian/pseudo--Riemannian spacetime is
a second rank tensorial object. Therefore $v_{\mu;\nu}$ can be
split into it's symmetric traceless, trace and antisymmetric parts.
These three parts constitute the shear, expansion and
rotation of the geodesic flow. Along the families of flow lines, one
can therefore write down the differential equations for each of these
quantities. It turns out that these equations are coupled and
quite difficult to solve completely. However, for simplistic
scenarios, there do exist solutions which have been analysed in some
amount of detail.

In Riemannian geometry and  GR one does not quite solve these equations. For general,
possibly non-geodesic flows,
one has the Raychaudhuri identity given
by :

\begin{equation}
\nabla_{\alpha}\left (v^{\mu}\nabla_{\mu}v^{\beta}\right )
= \left (\nabla_{\alpha}v^{\mu}\right )
 \left (\nabla_{\mu}v^{\beta}\right ) + v^{\mu}\nabla_{\alpha}\nabla_{\mu}
v^{\beta}
\end{equation}

In the case of affinely parametrised geodesic curves,
the left hand side is zero, which, thereby, gives us
the Raychaudhuri equation. Utilising the split of $v_{\mu ; \nu}$
into shear, rotation and
expansion one obtains equations for each of these quantities.
As an example, we have the
equation for the expansion :

\begin{equation}
\frac{d\theta}{d\lambda} +\frac{1}{n-1}\theta^{2} +2\sigma^{2} -
2\omega^{2} = -R_{\mu\nu}\xi^{\mu}\xi^{\nu}
\end{equation}

which, along with the other equations for the shear
and rotation, comprise the full set of nonlinear,
coupled ODEs. $n$ here is the dimension of space (for Euclidean signature)
or spacetime (for Lorentzian signature).
The equation for the expansion is analysed in a couple of ways. Here we
state one of them (for details see {\cite{wald}},
{\cite{hawking}}). It is easy to see that a redefinition
of $\theta$ in terms of the quantity $\theta=(n-1)\frac{F'}{F}$ converts
the equation into a second order linear ODE (note that the first order
equation was a nonlinear first order ODE, known in mathematics as a
Riccati equation). Thereafter, by imposing $\omega_{\mu\nu}=0$ (which is an
allowed solution of the evolution equation for rotation) we find that
the existence of zeros in a solution crucially depends on the
positivity (or nullity)  of the quantity $R_{\mu\nu}\xi^{\mu}\xi^{\nu}$.
Translating $R_{\mu\nu}$ into $T_{\mu\nu}-\frac{1}{2}Tg_{\mu\nu}$  via the
Einstein field equation one obtains a condition on the matter stress energy
known as the `strong energy condition' (this condition is the
statement that $T_{\mu\nu}-\frac{1}{2}Tg_{\mu\nu} \ge 0$.
This (and some weaker versions)
is, therefore, a requirement on the existence of zeros in F and
therefore an infinity in $\theta$. Since $\theta$ is related to the
rate of change of the cross--sectional area of the family of geodesics
(the cross-section being normal to the tangent vector field)
one finds that an infinity in $\theta$ (more precisely a negative infinity)
at a finite value of the affine parameter,
implies a vanishing of this area and therefore an intersection of the
set of geodesics. Such an intersection is termed focusing of geodesics
and is a precursor to the existence of spacetime singularities in the
sense of geodesic incompleteness, which, in some cases, does coincide with
the other definition of singularities--i.e. divergent curvatures.

Having stated the notion of focusing in the context of spacetime
geodesics we now return to quantum field theory. The RG flows
are integral curves of the
$\beta$-function in theory space,
with the added fact that they obey the RG equation. The curves may be
geodesic or non--geodesic. If they are geodesic then too they may
be affinely or non-affinely parametrised.
The overall analysis of the previous paragraphs
can therefore be translated to the case of RG flows
which may or may not be geodesic.

It has been shown by Dolan {\cite{dolan5}} that the RG equation
for a class of theories (exceptions and more general discussion with
an example are available in {\cite{broritz}})
can indeed be written as a condition on the $\beta$--function.
More precisely, $\beta^{a}$ (we use Latin indices
for theory space) is a conformal Killing vector. As mentioned
before, this was
previously assumed by Lassig {\cite{lassig}} in order to demonstrate that the
RG equation can be written (in theory space) in terms of a Lie derivative
of the
Green function with
respect to $\beta^{a}$.

This condition is :

\begin{equation}
\nabla_{a}\beta_{b} + \nabla_{b}\beta_{a} = - Dg_{ab}
\end{equation}

where $g_{ab} = \int \langle\Phi_{a}(y)\Phi_{b}(0)\rangle d^{D}y$
is the O'Connor--Stephens metric and $D$ is the dimension of Euclidean
space in which the field theory is defined. $\Phi_{a}$ are, in general,
composite operators and are defined after taking care of a
subtraction of $\langle\Phi_{a}\rangle$ (for details on this see {\cite{dolan3}}).
The metric in theory space is essentially the Fisher-Rao metric
(based on the Fisher information matrix--for details see{\cite{frieden}})
 used in statistics to compare probability distributions).

The above equation follows in a straightforward way, from the RG equation for
the two--point function :

\begin{eqnarray}
\kappa\partial_{\kappa} \langle\Phi_{a}(x)\Phi_{b}(y)\rangle +
\beta^{c}\partial_{c} \langle\Phi_{a}(x)\Phi_{b}(y)\rangle +
\partial_{a}\beta^{c} \langle\Phi_{c}(x)\Phi_{b}(y)\rangle + \\ \nonumber
\partial_{b}\beta^{c} \langle\Phi_{a}(x)\Phi_{c}(y)\rangle
=0
\end{eqnarray}

Using the facts that couplings are scaled to be dimensionless,
the $\Phi_{a}(x)$ have canonical mass dimension $D$ and the
scaling argument one can arrive at the equation :

\begin{eqnarray}
\left (x^{\mu}\frac{\partial}{\partial x^{\mu}} +
y^{\mu}\frac{\partial}{\partial y^{\mu}} \right )\langle\Phi_{a}(x)
\Phi_{b}(y)\rangle +2D \langle\Phi_{a}(x)\Phi_{b}(y)\rangle = \\ \nonumber
-\beta^{c}\partial_{c} \langle\Phi_{a}(x)\Phi_{b}(y)\rangle -
\partial_{a}\beta^{c} \langle \Phi_{c}(x)\Phi_{b}(y)\rangle -
\partial_{b}\beta^{c} \langle \Phi_{a}(x)\Phi_{c}(y)\rangle
\end{eqnarray}

Integrating over all $y$ and using translational invariance we get

\begin{equation}
\beta^{c}\partial_{c}g_{ab} +(\partial_{a}\beta^{c})
g_{cb} + (\partial_{b}
\beta^{c} )g_{ca} = -Dg_{ab}
\end{equation}

which will finally yield (after some simple manipulations)
the condition given earlier.

In this condition, if we use the following decomposition
:
\begin{equation}
\nabla_{a}\hat \beta_{b} = \sigma_{ab} + \omega_{ab} + \frac{1}{n-1}h_{ab}\theta
\end{equation}

where $\hat \beta^{a} = \frac{\beta^{a}}{\sqrt{\beta^{a}\beta_{a}}}$,
$h_{ab} = g_{ab}-{\hat \beta_{a}}{\hat \beta_{b}}$ and the
quantities $\sigma_{ab} $ (shear), $\omega_{ab}$ (rotation)
 and $\theta$ (expansion) are functions
of $\lambda$. Taking the trace of the above expression
w.r.t. $h^{ab}$  we arrive at the rather striking result that :

\begin{equation}
\theta = -\frac{D(n-1)}{2\beta}
\end{equation}

This is a general result. It states that RG flows must
necessarily converge ($\theta$ negative). They will converge
to a point where the norm of the $\beta$ function is zero, which
means (since the norm here is a positive quantity and we are
working in a Euclidean space) that the flows {\em focus where the
field theory has a fixed point} (i.e. all the $\beta$--functions
--each component of the vector field $\beta^{a}$ tend to zero).

The shear is given by the symmetric, traceless part of the
gradient of the normalised $\beta$--function.
It turns out to be :

\begin{equation}
\sigma_{ab} = \frac{1}{2} \left ( \nabla_{a}{\hat \beta_{b}}
+\nabla_{b}{\hat \beta_{a}}\right ) + h_{ab}\frac{D}{2\beta}
\end{equation}

In a similar way, the antisymmetric part leads to the rotation
, given through the expression :

\begin{equation}
\omega_{ab} = \frac{1}{2}\left ( \nabla_{a}{\hat \beta_{b}} -
\nabla_{b}{\hat \beta_{a}} \right )
\end{equation}

It is also possible that along certain directions in theory
space the $\beta$ functions are zero but they remain non--zero
along others. For example, in a three dimensional theory space
the geodesics which lie entirely on a two dimensional section
may converge to a point where the $\beta$--functions along the
coordinate directions which define this section, vanish.
If on has parallel sections where geodesics focus, then the
expansion does not tend to negative infinity--it is negative
nevertheless but we get a curve along which the convergence
happens. One can relate this to shear and rotation which are
nonzero if even one of the $\beta$--functions is finite.

Knowing the $\beta$ function and the metric in the coupling space
, one may therefore calculate these quantities. We shall do this
in detail later when we discuss a specific example.

The condition that $\beta^{a}$ be a conformal Killing vector
translates into the following under contraction with $\beta^{b}$. We have :

\begin{equation}
{\hat\beta}^{b}\nabla_{b}\beta^{a} = -\nabla^{a} \beta - D{\hat\beta}^{a}
\end{equation}

This equation
is a consequence derived from the RG equation.

The evolution equations for the quantities $\theta$, $\sigma_{ab}$
and $\omega_{ab}$ are the usual Raychaudhuri equations. However,
the major difference from ordinary geodesic flows in geometry
and RG flows in theory space is the fact that RG flows obey an
extra condition--notably that the $\beta$--function vector field
is conformally Killing. This determines $\theta$ universally for
all RG flows. Let us now write down the Raychaudhuri equation for the
expansion in terms of the $\beta$--function. This turns out to be
(for $\sigma_{ab}=0$, $\omega_{ab}=0$ which, are, consistent
solutions of the equations for $\sigma_{ab}$ and $\omega_{ab}$ :

\begin{equation}
\frac{D(n-1)}{2}\left [ \frac{d\beta}{d\lambda} + \frac{D}{2} \right ]
= -R_{ab}\beta^{a}\beta^{b}
\end{equation}

Note that this is a relation between the curvature properties of
theory space and the $\beta$--function. For geodesic flows both the
conformal Killing vector condition and the above equation has to
hold good simultaneously.

One can comment qualitatively on the realtion between curvature
properties of theory space and the existence of zeros of the
$\beta$--function. A divergence in $R_{ab}$ must be killed by
a zero in $\beta^{a}$ because the L. H. S. of the Raychaudhuri
equation if finite. Thus it is possible that the existence of
a curvature singularity in theory space indicates a fixed point of
the underlying field theory.

In a coupling space which is Einstein, i.e. $R_{ab} = \Lambda g_{ab}$
the above equation takes the form :

\begin{equation}
\frac{d\beta}{d\lambda} + \frac{D}{2}
= -C \beta^{2}
\end{equation}

where $ C = \frac{2\Lambda}{D(n-1)} $. The equation can easily be solved
to obtain the norm $\beta$, of the $\beta$--function vector field in terms
of the affine parameter. We have two cases to deal with $\Lambda > 0$
and $\Lambda <0$. The expressions for $\beta$ are :

\begin{equation}
\Lambda > 0 \vspace{.2in}; \vspace{.2in}
\beta = \sqrt{\frac{D}{2C}}
tan\sqrt{\frac{D}{2C}}\left ( C_1-C\lambda\right )
\end{equation}
\begin{equation}
\Lambda < 0 \vspace{.2in}; \vspace{.2in}
\beta = -\sqrt{\frac{D}{2\vert C\vert }}
tanh\sqrt{\frac{D}{2\vert C\vert }}\left ( C_1+\vert C\vert \lambda\right )
\end{equation}

where $ C= \frac{2\Lambda}{D(n-1)}$ and $C_1$ is an arbitrary constant.

The $\Lambda >0$ expression is defined over the interval $0\leq \left
(C_1 - C\lambda\right ) \leq \frac{\pi}{2}$, or between $\pi$ and
$\frac{3\pi}{2}$ and so on. This is to ensure non--negativity
of the norm of the $\beta$ function vector field. Similarly, for the
$\Lambda <0$ expression the interval is $-\infty \leq \left ( \vert C\vert
\lambda + C_1 \right ) \leq 0$. The focal points of the RG flows are
those where $\beta =0$ and these are seen to occur at finite values
of the parameter $\lambda$.

For a theory space of zero Ricci curvature, it is easy to see that
$\beta$ is linearly related to the parameter $\lambda$.

It is worth noting that these expressions for $\beta$ are conclusions
which are theory independent --it is possible that there could be
many theories for which the theory space is Einstein. The existence of
zeros at finite values of $\Lambda$ also follow largely from the
expressions for $\beta$--these correspond to the focal points of these
RG flows.

The equations for $\sigma_{ab}$ and $\omega_{ab}$ are the same as the
usual Raychaudhuri equations except for the fact that the expression
for $\theta$ needs to be substituted. As in the usual case $\sigma_{ab}$,
$\omega_{ab}$ both equal to zero are valid solutions of these equations.

We shall now illustrate the above results by working out
the details for the simplest of field theories--a
theory involving a single scalar with a $j\phi$ coupling (where $j$ is
a constant). The action is :

\begin{equation}
S(\phi;j,m^{2}) = \int d^{D}x \left (\frac{1}{2}\phi
\left (-{\nabla} ^{2} +m^{2} \right ) \phi + j\phi \right )
\end{equation}

From the action one can find out the corresponding $w(j,m^{2})$
($W=\int d^{D}x w$ and $Z=\exp (-W)$) after integrating over $\phi$.
Subsequently, the line element (for a Lagrangian linear in couplings the metric
$g_{ab}=-\partial_{a}\partial_{b}w$ \cite{dolan4}) is
given as :

\begin{equation}
ds^{2} = \left (-\partial_{a}\partial_{b}w \right )dx^{a}dx^{b} =
m^{2}d\xi^{2} + \frac{1}{2} \frac{1}{(4\pi)^{\frac{D}{2}}}{m^{2}}^{\frac{D}{2}
-2 } \Gamma (2-\frac{D}{2}) (dm^{2})^{2}
\end{equation}

The
mass--squared $m^{2}$ and the quantity  $\xi = <\xi> = -\frac{j}{m^{2}}$ are the two `couplings' (in
a generalised sense the mass is also attributed
the status of a coupling). The theory space here is
two--dimensional and we shall assume the Euclidean
coordinate space in which the theory is defined as
a flat Euclidean space of dimension $D$.

The metric given above can be transformed into the form :
\begin{equation}
ds^{2} = dr^{2} + r^{\frac{4}{D}}d\chi^{2}
\end{equation}

where :

\begin{eqnarray}
\chi = 2\sqrt{\pi}\left [ \frac{D}{4}\sqrt{\frac{2}{\Gamma(2-\frac{D}{2})}
}\right ] ^{\frac{2}{D}} \xi \\
r = \frac{4}{D} \sqrt{\frac{\Gamma(2-\frac{D}{2})}{2}}\left (\frac{m^{2}}{4\pi}
\right ) ^{\frac{D}{4}}
\end{eqnarray}

The  Ricci scalar is given as ;

\begin{equation}
R= -\frac{2(2-D)}{D^{2}r^{2}}
\end{equation}

Notice that for $D=2$ the Ricci scalar is zero and the space is also
flat (a coordinate transformation will reduce the polar coordinate
form for $D=2$ to $ds^{2} = dx^{2} + dy^{2}$).
The geometry for $D\neq 2$ is, however, nontrivial.  For $D>2$
can easily show that the line elements can be written
in the
form $ds^{2} = dr^{2} + a^2{(r)}d\chi^{2}$ (analgous to an Euclidean
, two--dimensional cosmological line element)

The $\beta$--functions can be obtained by using the scaling
properties of the couplings w.r.t a parameter, the values of
which define the renormalisation point. These turn out to be :

\begin{equation}
\beta^{r} = -\frac{D}{2}r \hspace{.1in} ; \hspace{.1in}
\beta^{\chi} = -\left (\frac{D}{2}-1 \right ) \chi
\end{equation}

The norm of the $\beta$ function in the coupling space
, given by $g_{ab}\beta^{a}\beta^{b}$ provides us with
the `normalised' $\beta$--functions which we use while
defining the expansion. It has been shown that for $D=2$
the flows are geodesic. The trajectories correspond to
radial curves which focus at the origin. The expansion
of the geodesic congruences (for $D=2$) also indicate such a behaviour :

\begin{equation}
\theta = -\frac{1}{\beta} = -\frac{1}{r}
\end{equation}

which implies that as $r\rightarrow 0$, $\theta\rightarrow -\infty$
which is the focal point of the congruence.

Note also that these $\beta$--functions satisfy the conformal
Killing vector condition.

Solving them we can obtain the integral curves $r(\lambda)$
and $\chi(\lambda)$. The result should tally with the
Raychaudhuri equation for the expansion $\theta$ written
down above. One can check easily that it does.

It is also easy to see that the shear and rotation for these flows in
$D=2$ are both identically equal to zero.

Let us now briefly frame the notions mentioned above
for the non-linear sigma model which describes strings.
Recall that one of the basic statements of string theory has been
to view the metric appearing in the sigma model as a {\em coupling}.
The dynamic geometry of
GR is a {\em derived} concept in string theory.
It is easy to see that if the metric is a coupling,
then the space of metrics is the corresponding theory space.
Therefore, the Raychaudhuri equation
can be used to analyse the nature of the RG flows in this
`superspace'. For simplicity, let us qualitatively look at
bosonic strings. Here the metric $g_{\mu\nu}$, the anti--symmetric
tensor potential ($B_{\mu\nu}$) and the dilaton $\phi$ are the
background fields
to which the string couples. The full theory space $\left ( {\bf g, B, \phi
}\right )$ is therefore a space of functions. The $\beta$--functionals
for each of these `couplings' are known from the RG analysis of the
nonlinear $\sigma$ model. As a simple example, one may
think of a minisuperspace model with only a single function
$a(t)$ specifying the metric and a dilaton $\phi(t)$ as a second
`coupling'. The space of theories is therefore the
space of $a(t)$, $\phi(t)$--a function space. It is therefore
necessary to evaluate the metric in this space which is
obtainable from the two-point functions of the nonlinear
sigma model. Once we know the metric, we can draw
conclusions on geodesic flows,
curvature properties of theory space and the Raychaudhuri identity.
An analysis somewhat in the same vein was carried out in
{\cite{tseyt:hepth}} where attractive and repulsive behaviour
of the RG trajectories in theory space (with a minisuperspace
ansatz) was discussed.

In models of particle physics the evolution of couplings, the
nature and extrapolation  of RG flows and their intersection do play an
important role in understanding the unification of interactions
It is certainly worth investigating whether the
methods and results presented here can be of any use there.

Let us conclude by summarising the main results :

\begin{itemize}
\item{It has been shown that the expansion for an RG flow
is a negative quantity depending directly on coordinate and theory
space dimensions and inversely on the norm of the $\beta$--function
in theory space. The shear and rotation for these flows may/may not be zero.
Both
these facts result from the geometric version of the RG equation.}
\item{The Raychaudhuri equation for the expansion has been rewritten in
terms of the $\beta$-function and it serves as an additional constraint
on the $\beta$--function. RG flows are generated by conformally Killing
vector fields and their evolution (for geodesic flows) is indeed
governed by the Raychaudhuri equations.}
\end{itemize}

The salient feature of the  analysis presented here is, that
methods and conclusions obtained in coordinate space in
Riemannian geometry and GR can be used (with a few appropriate
modifications)
to analyse the
the geometry  and physics in the space of couplings of a given
quantum field theory. It is not clear as yet whether this
approach helps in some way in understanding quantum field theory
or it is just an exercise which will remain an exclusively
aesthetic endeavour. However, we should note that
the results on the nature of RG flows in
theory space do not pertain to
any {\em specific } field theory but are applicable to the
class of field theories for which the $\beta$--function
satisfies the conformal Killing condition. For theories where
the conformal Killing vector character of the $\beta$--function
does not hold (such as the one dimensional Ising model as shown
in {\cite{broritz}}) one will have to write down the
full set of Raychaudhuri equations (without assuming specific
forms of $\sigma_{ab}$ or $\theta$ ) for geodesic RG flows
. Then one may proceed to further the analysis. A crucial
assumption in all this, is the metric ansatz which,
however, has been used extensively in this context and follows
from statistics.
Additionally, the fact that the theory space in the QFT of strings
coincides with the actual space of metrics is a problem
worth pursuing. So is the application in realistic models
of particle physics.
Further detailed calculations along the each of these directions, as well as
more nontrivial examples illustrating the ideas presented here,
will be communicated elsewhere.

\end{document}